# Bifurcation of Stretched Exponential Relaxation in Microscopically Homogeneous Glasses

G. G. Naumis[1] and J. C. Phillips[2]

1. Univ Nacl. Autonoma Mexico, Inst Fis, Mexico City 01000, DF Mexico

2. Dept. of Physics and Astronomy, Rutgers University, Piscataway, N. J., 08854-8019

Abstract

Measured exponents associated with Stretched Exponential Relaxation (SER) are widely scattered in microscopically inhomogeneous glasses, but accurately bifurcate into two "magic" values, 3/5 and 3/7, in a wide variety of microscopically homogeneous glasses. These bifurcated values are derived here from a statistical product model that involves diffusion of excitations to native traps in the presence of short-range forces only, or combined short- and long- range forces, respectively. Bifurcated SER can be used to monitor sample homogeneity. It explains a wide range of experimental data, and even includes multiple aspects of the citation distributions of $20^{th}$ century science, involving 25 million papers and 600 million citations, and why these changed radically in 1960. It also shows that the distribution of country population sizes has compacted glassy character, and is strongly influenced by migration.

\body

Stretched exponential relaxation (SER) was first recognized by Kohlrausch in 1854 as providing a strikingly economical and accurate fit (for example, far superior to two exponentials, although the latter involve 4 parameters) to the residual glassy decay of charge on a Leiden jar, and modern data have increasingly confirmed this superiority for relaxation of glassy excitations, up to as many as 12 decades in time for g-Se (1). The stretched exponential function $A\exp(-(t/\tau)^\beta)$ contains only 3 parameters, with $\tau$ being a material-sensitive parameter and $0 < \beta < 1$ behaving as a kinetic fractional exponent. The function itself is not analytic (like power laws and exponentials; instead, it belongs to the limbo of asymptotic functions). There have been many different theoretical models of SER (1), and for this reason some regard SER as the oldest unsolved problem in science. Broadly speaking, most of these models regard both $\tau$ (material-dependent) and $\beta$ (dimensionless) merely as adjustable parameters, with $\beta$ subject only to weak conditions (such as (2,3) $1/3 \leq \beta \leq 1$).

**Experimental Data Base**

The experimental situation depends on the materials studied, as SER-like "fat tails" have been observed in many sparse distributions, such as radio and light emissions from galaxies, of US GOM OCS oilfield reserve sizes, of World, US and French agglomeration sizes, of country population sizes, of daily Forex US-Mark and Franc-Mark price variations, of Vostok (near the south pole) temperature variations over the last 400 000 years, of the Raup-Sepkoski's kill curve, and earthquake sizes and fault displacements (4). Here we focus on the rich data available for ideal glasses, by which we mean modern microscopically homogeneous glasses (1,5,6,7). Most theories assume microscopic homogeneity, yet unless very special care in sample preparation has been taken, glass samples are microscopically inhomogeneous on micron length scales. This was surely the case for the uncontrolled examples found in older comprehensive but uncritical compilations that included polymer blends, resins and other samples of dubious homogeneity dating back many decades (8). Included in the rich "ideal" microscopically homogeneous data base are network and molecular glasses, some polymers, electronic glasses, and a few large-scale numerical simulations that explored relaxation over at least 2-3 decades for non-crystallizing models, for example bidispersive hard spheres, compared to as little as one decade or less for microscopically inhomogeneous glasses.

Each of the 50 entries in the curated "rich" glass data base has been carefully examined to identify multiple concordances of β with other entries with respect to glassy material, pump and probe, in time-resolved, field-free studies (dielectric relaxation experiments are excluded, for many reasons (1,5,6,7)). In these critically studied ideal β examples SER typically spans three or more decades in time; an executive summary of curated values is shown elsewhere (6) as table I. In Q-dependent spin-polarized neutron diffraction experiments on molecular glasses relaxing quasi-particles are identified with momentum transfers $Q = Q_1$, where a large and narrow peak in the intermediate scattering function $S(Q)$ occurs at $Q = Q_1$. This peak can be identified with diffusing quasi-particle excitations in simple polymer glasses like polybutalene (PB).

In contrast to collecting all data and indiscriminately giving each entry equal weight, curating a large data base that is still dominated by entries primarily from inhomogeneous (uncontrolled) samples is a complex and even unpopular task that requires uncommon sense and must be handled with care; the interested reader can find extensive discussions of "β curation" elsewhere (1,5,6,7). Fig. 1 shows a canonical example of both the selective and predictive power of a β theory based on curated multiple concordances. There are many such examples (1,5,6,7), but synthetic rubber (PB) is especially dramatic. Here the 1994 theoretical prediction based on multiple concordances (a) unhesitatingly rejected the previously published "bad" NRL data (strongly % vinyl-dependent), and (b) successfully predicted the constant value "good" result (nearly % vinyl-independent) obtained at Grenoble by spin-polarized neutron spectroscopy (the ns non-contact gold standard), two years later!

**Results**

The entries in the curated data bases reported in the 1996, 2006, and 2010 reviews span a wide range of materials and properties, yet they show a surprisingly simple universal pattern, much different from earlier indiscriminate and broadly scattered patterns (8): for glasses dominated by short-range Brownian forces only, $\beta = 0.60 \pm 0.01$, while for polymers (long-range intrachain elastic forces) and cases where Coulomb (also long-range) forces were present, $\beta = 0.43 \pm 0.01$. The crossover from 3/5 to 3/7 occurs rarely, but it has been observed in the case of monomer (PG) to polymer (PPG) propyl glycol (5,6). It may be that these fractal values of $\beta = 3/5$ and $3/7$ are the only two "magic numbers" known in science outside of quantum phenomena. These values may be technologically useful for testing the quality of glass samples on nm to m length scales, and they can also be used to identify glass-like effects of compacted network self-organization in unexpected contexts.

**Derivation**

Fortunately, older theoretical models, including a simple spherical trap capture model (1,9), had already predicted $\beta = 3/5$ before the modern ideal data became available, leaving only the mixed case of short- and long-range forces with $\beta = 3/7$ as an unexplained puzzle. Here an extension of the spherical capture model explains the mixed ideal case in terms of anomalous superdiffusion (or Lévy flights). In the old model (9) excitations diffuse until they are captured by randomly distributed immobile glass traps. Because these traps are randomly distributed, they correspond to a microscopically homogeneous glass, yet their presence makes the glass nanoscopically inhomogeneous. The result of the nanoscopic diffusion-to-traps model is based essentially on scaling by dimensional analysis of the diffusion equation and is topological:

$$\beta = d/(d+2) \tag{1}$$

where d is the spatial dimensionality $[(1/3 \leq \beta \leq 1) \leftrightarrow (1 \leq d \leq \infty)]$. Topological relations are independent of geometry, so the model replaces the nearest trap by a spherical trap shell, after which the resulting calculation is straightforward for short-range forces only (9).

Suppose that the diffusive motion is instead determined by competition between Brownian short (**r**) - and Lévy long-range (**R**) forces, and that the respective length scales are $l$ and L. Also suppose that the average distance between immobile traps is $\lambda$ (also the radius of the spherical capture construct (9)), so that there are two nanoscopically distinguishable length scales

$$l < \lambda < L \tag{2}$$

In this intermediate case the diffusive motion is separated into two kinds of steps, Brownian and Lévy, it becomes entangled, and the simple short-range solution $\beta_s = d/(d+2)$ is no longer valid. To see what happens in the entangled case, consider the following example (10) from polymer science. Dense packing of polymer glasses tends to favor polymer bundles, with individual polymer chains having variable lengths ~ L. Here excitations diffuse to polymer chain ends, which act as traps, but not all the chain ends are at bundle ends; many chains are broken within bundles (length scale $\lambda$). Diffusing excitations can reach these internal traps by diffusing with large steps to chain ends (length scale L) bypassing many adjacent traps, or by hopping between chains (short steps, length scale $l$) to reach staggered traps on nearby chains. This entangled process implies composite relaxation paths that combine small Brownian interchain and large Lévy intrachain steps (10).

Products of discrete probability distributions often occur in connection with stretched exponential-like frequency-rank functions (11), so we analyze composite relaxation paths by dividing the excited density ρ(t) harmonically, as

$$\rho(t) = \rho_s(t/\tau_s)\rho_l(t/\tau_l) \qquad (3)$$

The local diffusion equation in normal liquids is based on short-range Brownian **r** forces and is written

$$\partial\rho/\partial t = D\nabla^2_{\mathbf{r}}\rho \qquad (4)$$

In glasses with short-range **r** forces the diffusion to traps model leads through scaling by dimensional analysis (9) to (1), $\beta_s = d/(d+2)$, which is in excellent agreement with earlier results (1-4) and those discussed here. How should Eqn. (4) be generalized to describe the nonlocal case of mixed short Brownian **r**- and long Lévy **R**- range forces? The physical content of Eqn. (4) is that the spatial fluctuations on an **r** scale drive the temporal diffusion. In the mixed **r:R** case, short-(long-)range forces drive spatial fluctuations on an **r** (**R**)scale, with the two processes (with different length scales being separated by the trap spacing) competing to drive temporal diffusion. Such nonlocal competition between short- and long-range forces (and small and large (Lévy flight) diffusive steps) suggests the anomalous diffusion equation with superposed fluctuation channels

$$\partial\rho/\partial t = (D_1\nabla^2_{\mathbf{R}} + D_2\nabla^2_{\mathbf{r}})\rho \qquad (5)$$

The superposition approximation from Eqn. (4) to Eqn. (5) has been studied and validated by extensive numerical simulations in the context of anomalous diffusion probe trajectories described as continuous random walks impacted by bath statistics (12). Substituting Eqn. (3) in Eqn. (5) causes the left hand side to separate into two terms for $\partial \rho_s/\partial t$ and $\partial \rho_l/\partial t$. The right hand side also separates, and the optimal combination of relaxation by combined short- and long-range forces is determined by the principle of maximum entropy production (13). One can assume that at the variational extremum this entropic optimization produces a separation of the right hand side of Eqn. (5) fully parallel to that of the left hand side. This leads to decoupled equations for $\rho_l(\mathbf{R})$ and $\rho_s(\mathbf{r})$, each equation formally identical to Eqn. (4). With $\rho_l(\mathbf{R})$ scaling as $L^{-fd}$, and with $\rho_s(\mathbf{r})$ scaling as $L^{-(1-f)d}$, Eqn. (5) can then be solved by scaling by dimensional analysis (9) twice to the separately local $\mathbf{r}$ and $\mathbf{R}$ equations to give

$$I(t) \sim \exp[-(t/\tau_1)^{\beta_l}] \exp[-(t/\tau_2)^{\beta_s}] \qquad (6)$$

with $\beta_l = fd/(fd + 2)$ and $\beta_s = (1-f)d/((1-f)d + 2)$. The fastest relaxation entropy production (11) at long times will occur along paths for which $f = 1/2$ and $\beta_l = \beta_s = d/(d+4)$, or 3/7 for d = 3. Thus the product assumption (3) explains the overwhelmingly reproducible appearances of $\beta$ = 3/5 or 3/7 in ~ 50 examples of microscopically homogeneous glasses (6), emphasized in the abstract of (1).

**Two Examples**

Here we discuss specifically only two examples of β bifurcation due to long-range forces, but many other examples, including both experiments and numerical simulations, are given in (6). The extra "push" given to short-range Brownian fluctuations by long-range Lévy forces explains the anomalously enhanced (by a factor of 100!) surface diffusion reported for fragile molecular ideal glass formers like OrthoTerPhenyl (14).   It is striking that while the measured diffusivity is enhanced by long-range interactions, its long time-dependent relaxation is slowed, as β is reduced by entanglement. Analysis of photobleached OTP relaxation data also shows a dramatic bifurcation into β = 3/5 and 3/7, with β = 3/7 corresponding to native OTP molecules and similarly tricyclic anthracene impurities, while five other dissimilar (not tricyclic) molecules belonged to a β = 3/5 branch.  These results are shown in Fig. 2; details are elsewhere (6). The key point is that knowledge of the relation (1), and specifically the bifurcated magic fractions 3/5 and 3/7, leads immediately to a natural interpretation of the original data, which is not possible otherwise.

There are many precise ideal physical examples (1,5,6,7) of SER in the presence of competing short- and long-range forces.  Elegant recent studies of decay of luminescence from 16 single-crystal isoelectronic $ZnSe_{1-x}Te_x$ alloys of commercial quality (these microscopically homogeneous alloys are used in orange light-emitting diodes) (15) exhibited SER with a fitting accuracy of 0.2%.  Two previously unnoticed extrema in τ(x) or β(x) are associated (6) with β = 3/5 or 3/7.  At x = 0 and 1, geminate recombination with β = 1 occurs, so the crossover in these alloys from quantum order to classical glassy disorder is a very interesting future problem. Simulations for a physically apparently different problem (lipid bilayers) gave 0.45(0.60) for atomistic (coarse-grained) models without and with Coulomb cutoffs (16); these leave little doubt of the origin of the microscopic origin of β bifurcation.  An historical note: Eqn. (4) is well-known for liquids (17), while Eqns. (3,5,6) seem to be new and specific to microscopically homogeneous glasses.

**Web of Science**

A surprising feature of the 3/5, 3/7 SER bifurcation discussed here is that it has recently appeared in a completely unexpected context, the statistics of citation chains (18). As in many statistical problems, the distributions of sparse citation statistics can be fitted in more than one way, but the fits become unambiguous for large ("rich") data bases. The comprehensive study of 20th century citations, involving 25 million papers and 600 million citations (18), with $10^6 - 10^7$ more entries than typical sparse data bases fitted by power-laws, and unique in the history of epistemology, greatly expanded the previous data base of "only" a few million citations (19). In this ultralarge data base nonanalytic stretched exponentials (SE) give unambiguous best fits to citation chains for both low and intermediate citation levels $n < n_1$, where the crossover point $n_1$ ~ 40 earlier in the century, and $n_1$ ~ 200 later. The middle SE region accounts for 95% of the citations, and it, not the scale-free or power law 5% high end, probably objectively represents the essential features of working research citation patterns. The 5% power law high end is explained by cumulative advantage (fame, or "the rich get richer"), and has long been known from linguistic studies of word frequencies (20).

This 2009 study of 25 million papers and 600 million citations found bifurcated SE exponents β close to 3/5 and 3/7 which describe citation chains before and after 1960, respectively (see Fig. 3, included here for the reader's convenience, and also elsewhere (7) as Fig. 8). Within the framework of the short-long-range forces model discussed here, this transition is interpreted as reflecting the effects of globalization in 1960 on scientific research, as seen in the context of a microscopically homogeneous glassy scientific community. Globalization of science occurred abruptly in 1960 because of the confluence of several factors (7) favoring intercontinental conferences (the space race and jet travel).

More detailed reasons behind the unexpectedly close correspondence between scientific research, as reflected by citation chains, and glass networks are discussed extensively elsewhere (6,7). Note that most scientific research is carried out in the compacted context of peer-related scientific research, and radically new ideas ("outliers") are rare. Note also that the compactive analogy with glass networks is objective and physical, and makes multiple quantitative predictions accurate to a few %, although it contains no adjustable parameters. It is not based subjectively on axiomatic anthropomorphic individual or social assumptions, although these have often been used to explain the power-law high end (18). The appearance of long-range diffusive interactions in globalized research is reminiscent of long-range hopping in the Lévy-flight foraging hypothesis for superdiffusion (21). The importance of the condition of microscopic homogeneity is brought out by comparing the world-wide value of $\beta$ (>1960) = 0.47 with the more homogeneous data bases (1981-1997, ISI, $\beta \sim 0.39$; PRD, $\beta \sim 0.44$) in the $\sim 7$ million citation data base (22). As the > 1960 data bases became more homogeneous (but smaller), their average approached 0.43 (but with larger scatter).

**Other Models**

Subjective axiomatic anthropomorphic individual or social assumptions (23) tend to give simple exponential functions ($\beta = 1$) below $n_1$, which do not explain fat tail distributions (4), notably the 600 million citations discussed in (18). However, one does find simple exponentials in time delays of responses to email messages (24), an isolated and much simpler task than preparing interlocking networked citation lists for research papers. This difference can even be interpreted as an objective definition of the differences between the much used (but seldom quantified) objective/subjective dichotomy (in the abstracts of $\sim 10^3$ papers/year over the last 25 years).

Simple exponentials correspond to ballistic (random or noninteracting) spreading of excitations, while stretched exponentials correspond to diffusion with strong interactions. In the last 25 years, there have been 140 scientific abstracts discussing "diffusion of knowledge" (25), and none discussing "ballistics of (or random) knowledge". The common assumption of simple exponential distributions associated with random citations appears unjustified (23), as is the misleadingly small value of the crossover to fame $n_1$ (about 25 (23), instead of 200 (18)) which it suggests.

**Broad Implications**

It is striking that modern distributions of country population sizes (except China and India outliers) are fitted over nearly two decades (4) by $\beta = 0.42$; this implies extensive "long hops" [large scale migration]. It also implies that the world is not "empty" (like a gas, $\beta = 1$), nor is it even "full" (like a liquid, $\beta \sim 0.8$), it is a compacted "overfull" world that has solidified (like a glass, $\beta = 0.60$ or $0.43$) (26). It also suggests the existence of a broad class of homogeneously compacted "glassy" networks that includes both network glasses and proteins (27), as well as crowded citation networks and organized populations.

We are grateful to M. L. Wallace, Y. Gingras and L. J. Sham for helpful comments.

12

## Figure Captions

Fig. 1. Dependence of β on vinyl fraction in polybutadiene. The older 1991 data are compared (1) with 1996 spin-polarized neutron data. Note that in 1994 the concordant diffusion-to-traps theory contradicted the poor 1991 NRL data, and correctly predicted the Grenoble 1996 data.

Fig. 2. Relaxation of photobleaching in OTP exhibits two branches of β, bifurcated about 3/5 and 3/7 for small molecular volumes. Here defect points for Ice I and II have been added. For the molecular points the volume (upper abscissa) is normalized to that of OTP, while for the defect Ice points it is normalized to the volume of $H_2O$. As the normalized molecular volumes increase beyond 1, the dimensionality d increases beyond 3, and $\beta = d/(d+2)$ tends to 1. The figure shows how β in OTP can be used to establish a relative length scale for defects in Ice, starting from the data for β of heteromolecules in OTP, which is a remarkable demonstration of the power of SER analysis to transcend materials. See (6) for details.

Fig. 3. Evolution of the number of papers n with a number of citations τ (or T) as a function of the decade in which the paper was published. The distributions for the decades up to 1960 are well fitted by an SER with $\beta = 0.57$, while the decades after 1960 are fitted with $\beta = 0.47$. The abrupt crossover at 1960 is unambiguous (18). A glassy

model of citation chains is given elsewhere (6,7). It explains the small shifts away from 3/5 and 3/7 in terms of a matrix which preserves the sum of the two shifts.

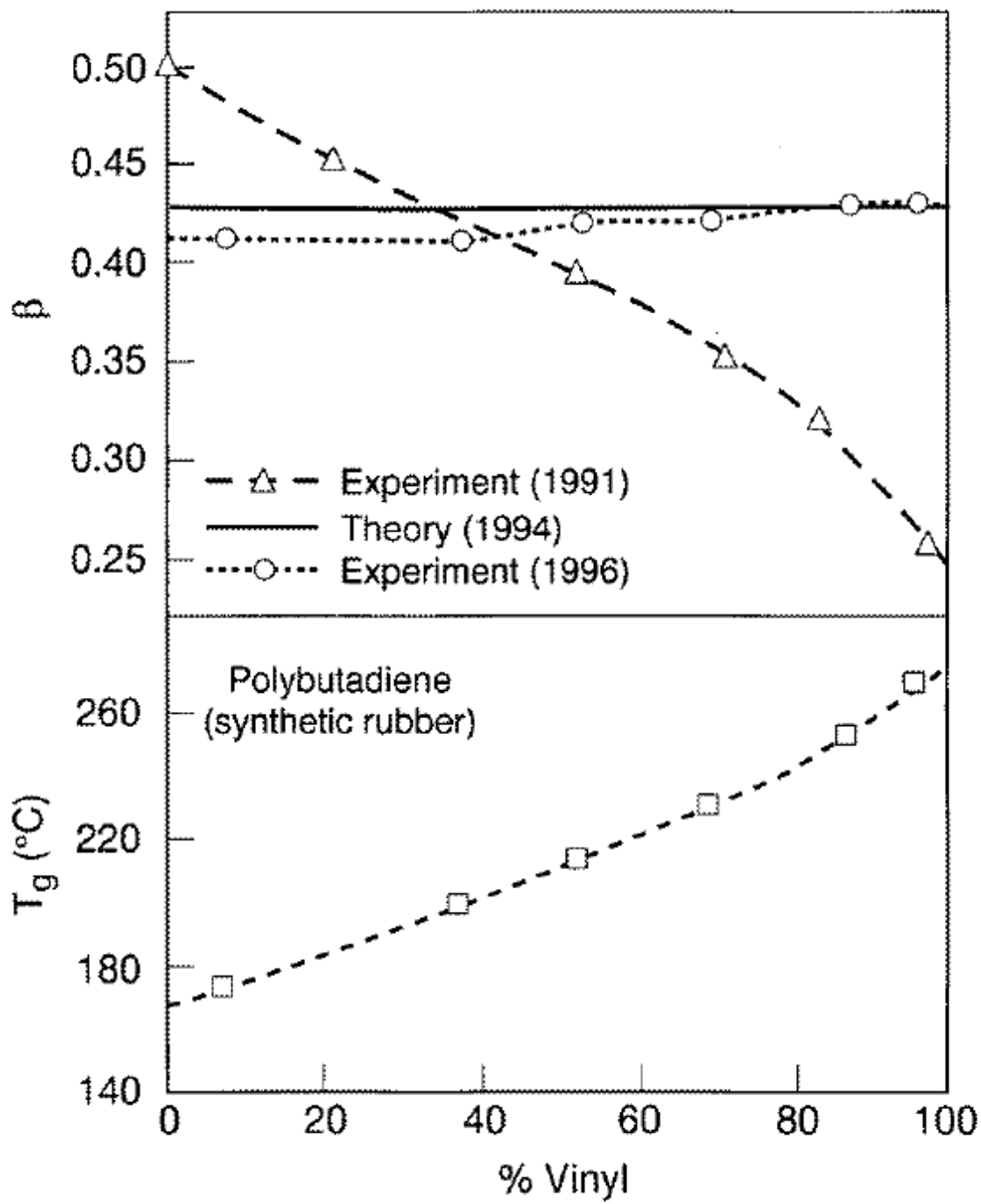

Fig. 1.

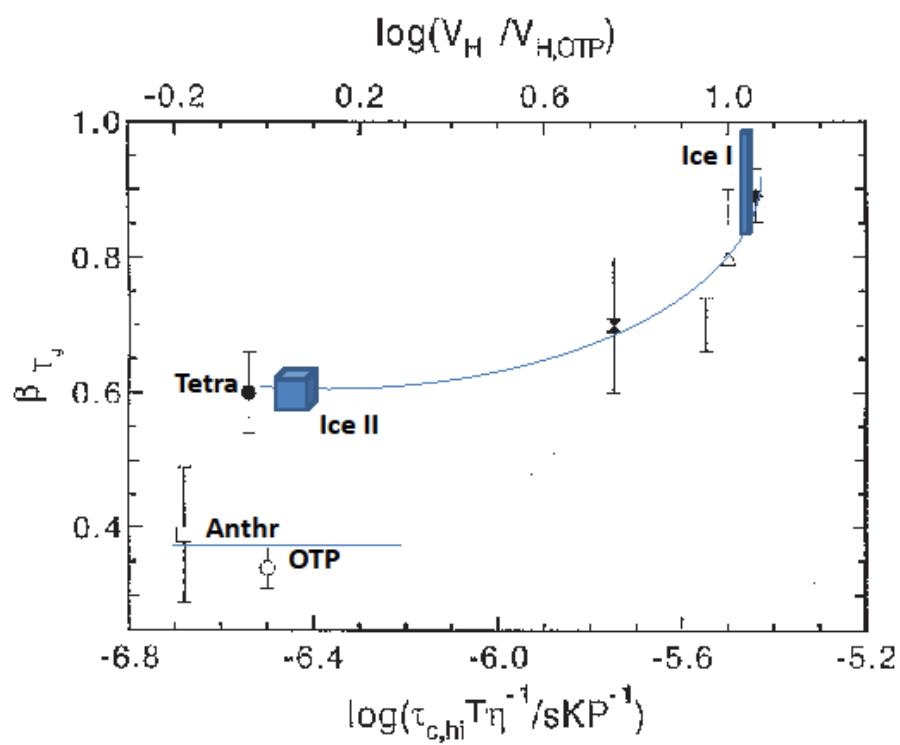

Fig. 2

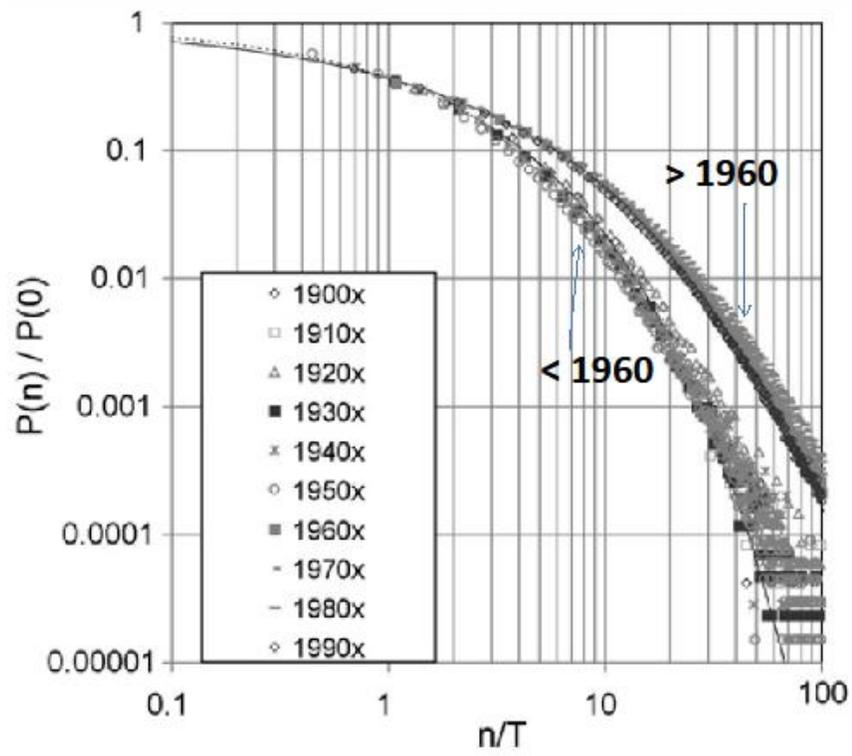

Fig. 3.